\documentclass{aa}
\usepackage{graphicx}
\usepackage{epsfig}

\def\r0{r_{0}}

\def\r{\bf r}

 \def \ie{{\rm i.e.}}

\def \ie{{\rm i.e.}}
\def \aa{A\&A}

\def \av{\ensuremath{A_V}}

\def\cc{\ifmmode{\,{\rm cm}^{-3}}\else{$\,{\rm cm}^{-3}$}\fi}
\def\cq{\ifmmode{\,{\rm cm}^{-2}}\else{$\,{\rm cm}^{-2}$}\fi}
\def\eccs{\ifmmode{\,{\rm erg}\,{\rm cm}^{-3} {\rm s}^{-1}}\else{$\,{\rm
erg}\,{\rm cm}^{-3} {\rm s}^{-1}$}\fi}
\def\deg{\ifmmode{^{\circ}}\else{$^{\circ}$}\fi} 
\def\pc{\ifmmode{\,{\rm pc}}\else{$\,{\rm pc}$}\fi} 
\def\kms{\ifmmode{\,{\rm km}\,{\rm s}^{-1}}\else{km s$^{-1}$}\fi} 
\def\kmspc{\ifmmode{\,{\rm km}\,{\rm s}^{-1}\,{\rm pc}^{-1}}\else{km
s$^{-1}$ pc$^{-1}$}\fi} 
\def\Kkms{\ifmmode{\,{\rm K\,km\,s}^{-1}}\else{$\,{\rm K\,km\,s}^{-1}$}\fi} 
\def \twCO{\ifmmode{\rm ^{12}CO}\else{$\rm^{12}CO$}\fi} 
\def \thCO{\ifmmode{\rm ^{13}CO}\else{$\rm^{13}CO$}\fi} 
\def \Cp{\ifmmode{\rm C^+}\else{$\rm C^+$}\fi} 
\def \Hp{\ifmmode{\rm H^+}\else{$\rm H^+$}\fi} 
\def \CHp{\ifmmode{\rm CH^+}\else{$\rm CH^+$}\fi}
\def \CHtp{\ifmmode{\rm CH_2^+}\else{$\rm CH_2^+$}\fi} 
\def \CtH{\ifmmode{\rm C_2H}\else{$\rm C_2H$}\fi} 
\def \CtHt{\ifmmode{\rm C_2H_2}\else{$\rm C_2H_2$}\fi} 
\def \CtHtp{\ifmmode{\rm C_2H_2^+}\else{$\rm C_2H_2^+$}\fi} 
\def \CtHp{\ifmmode{\rm C_2H^+}\else{$\rm C_2H^+$}\fi} 
\def \CHt{\ifmmode{\rm CH_2}\else{$\rm CH_2$}\fi} 
\def \CHthp{\ifmmode{\rm CH_3^+}\else{$\rm CH_3^+$}\fi} 
\def \Hthp{\ifmmode{\rm H_3^+}\else{$\rm H_3^+$}\fi} 
\def \HCOp{\ifmmode{\rm HCO^+}\else{$\rm HCO^+$}\fi} 
\def \HthOp{\ifmmode{\rm H_3O^+}\else{$\rm H_3O^+$}\fi} 
\def \HCfiN{\ifmmode{\rm HC_5N}\else{$\rm HC_5N$}\fi} 
\def \wat{\ifmmode{\rm H_2O}\else{$\rm H_2O$}\fi} 
\def \oxy{\ifmmode{\rm O_2}\else{$\rm O_2$}\fi} 
\def \HH{\ifmmode{\rm H_2}\else{$\rm H_2$}\fi}
\def \Jone{\ifmmode{\rm {(J=1--0)}}\else{{(J=1--0)}}\fi} 
\def \Jtwo{\ifmmode{\rm {(J=2--1)}}\else{{(J=2--1)}}\fi} 
\def \Jthr{\ifmmode{\rm {(J=3--2)}}\else{{(J=3--2)}}\fi} 
\def \Jfou{\ifmmode{\rm {(J=4--3)}}\else{{(J=4--3)}}\fi} 
\def \Jfo{\ifmmode{\rm {J=4--3}}\else{{J=4--3}}\fi} 
\def \Jon{\ifmmode{\rm {J=1--0}}\else{{J=1--0}}\fi} 
\def \Jtw{\ifmmode{\rm {J=2--1}}\else{{J=2--1}}\fi} 
\def \Jth{\ifmmode{\rm {J=3--2}}\else{{J=3--2}}\fi} 
\def \Jfi{\ifmmode{\rm {J=4--3}}\else{{J=4--3}}\fi} 
\def \Ta{\ifmmode{\rm T_A}\else{$\rm T_A$}\fi} 
\def \Tas{\ifmmode{T_{\rm A}^*}\else{$T_{\rm A}^*$}\fi} 
\def \Tmb{\ifmmode{\rm T_{mb}}\else{$\rm T_{mb}$}\fi} 
\def \Tr{\ifmmode{\rm T_r}\else{$\rm T_r$}\fi} 
\def \Trs{\ifmmode{\rm T_r^*}\else{$\rm T_r^*$}\fi}
\def \Tkin{\ifmmode{\rm T_{kin}}\else{$\rm T_{kin}$}\fi}
\def \eqref#1{eq.(\ref{#1})}

\begin{document}

\title{Dissipative structures of diffuse molecular gas:
\\ I -  Broad \HCOp\Jone\ emission.}

\author{E. Falgarone\inst{1}, G. Pineau des For{\^e}ts\inst{2},
P. Hily-Blant \inst{3}, P. Schilke \inst{4}}

\offprints{E. Falgarone}

\institute{LERMA/LRA, CNRS-UMR 8112, Ecole Normale
  Sup{\'e}rieure, 24 rue Lhomond, F-75005 Paris, France \and
  IAS, CNRS-UMR 8617, Universit\'e Paris-Sud, F-91405 Orsay,
  France, and LUTH, Observatoire de Paris, F-92195 Meudon,
  France \and IRAM, 300 rue de la Piscine F-38406 St. Martin
  d'H\`eres, France \and MPIfR, Auf den H\"ugel 69, D-53121
  Bonn, Germany}

\date{Received ../../05; accepted ../../..}

\abstract{Search for specific chemical signatures of intermittent
  dissipation of turbulence in diffuse molecular clouds.
We observed \HCOp(1-0) lines and the two lowest rotational
  transitions of \twCO\ and \thCO\ with an exceptional
  signal-to-noise ratio in the translucent environment of
  low mass dense cores, where turbulence dissipation is
  expected to take place.  Some of the observed positions
  belong to a new kind of small scale structures identified
  in CO(1-0) maps of these environments as the locus of
  non-Gaussian velocity shears in the statistics of their
  turbulent velocity field \ie\ singular regions generated
  by the intermittent dissipation of turbulence.
We report the detection of broad \HCOp(1-0) lines ($10 {\rm
  mK}< \Tas < 0.5$ K). The interpretation of 10 of the
  \HCOp\ velocity components is conducted in conjunction
  with that of the associated optically thin \thCO\
  emission.  The derived \HCOp\ column densities span a
  broad range, $10^{11}< N(\HCOp)/\Delta v <4 \times
  10^{12}$ \cq/\kms, and the inferred \HCOp\ abundances, $2
  \times 10^{-10}<X(\HCOp) < 10^{-8}$, are more than one
  order of magnitude above those produced by steady-state
  chemistry in gas weakly shielded from UV photons, even at
  large densities.  We compare our results with the
  predictions of non-equilibrium chemistry, swiftly
  triggered in bursts of turbulence dissipation and followed
  by a slow thermal and chemical relaxation phase, assumed
  isobaric.  The set of values derived from the
  observations, \ie\ large \HCOp\ abundances, temperatures
  in the range of 100--200 K and densities in the range
  100--10$^3$ \cc, unambiguously belongs to the relaxation
  phase. The kinematic properties of the gas suggest in turn
  that the observed \HCOp\ line emission results from a
  space-time average in the beam of the whole cycle followed
  by the gas and that the chemical enrichment is made at the
  expense of the non-thermal energy.  Last, we show that the
  "warm chemistry" signature (\ie\ large abundances of
  \HCOp, \CHp, \wat\ and OH) acquired by the gas within a
  few hundred years, the duration of the impulsive chemical
  enrichment, is kept over more than thousand years. During
  the relaxation phase, the \wat/OH abundance ratio stays
  close to the value measured in diffuse gas by the \textit{
  SWAS} satellite, while the OH/\HCOp\ ratio increases by
  more than one order of magnitude.}

\titlerunning{Broad \HCOp\Jone\ emission in dense core environments}
\authorrunning{Falgarone et al.}

\maketitle


\section{Introduction}

Parsec scale maps of molecular clouds with only moderate
star formation provide a comprehensive view of the
environment of dense cores before their disruption by young
star activity. Whether the tracer is extinction, molecular
lines or dust thermal emission, large scale maps reveal long
and massive filaments of gas and dust. Dense cores are most
often nested in these filaments, which are likely their site
of formation.  The open questions raised by star formation
thus include those on dense core and filament formation.
Filaments being by essence structures at least one order of
magnitude longer than thick, their observational study
requires maps with large dynamic range.  These observations,
now available, open new perspectives: filaments are
ubiquitous and span a broad range of mass per unit length
(e.g. \cite{aa94}, \cite{mizu95} and \cite{haik04} for
massive filaments and \cite{phb04}, \cite{fpw91},
\cite{fpp01} for most tenuous ones).

Recent observations of the environment of low mass dense
cores in \twCO\ and \thCO\ rotational transitions, have also
brought to light previously hidden velocity structures.  The
turbulent velocity field of core environments, alike
turbulent gas, exhibits more large velocity shears at small
scale than anticipated from a Gaussian distribution and the
spatial distribution of the non-Gaussian occurrences of
velocity shears delineates a new kind of small scale
structure in these core environments: their locus forms a
network of narrow filaments, not coinciding with those of
dense gas (\cite{pf03}, \cite{phb04}, \cite{hfp06},
hereafter Paper IV). It has been shown by \cite{lis96} that
they trace the extrema of the vorticity projected in the
plane of the sky.  As discussed in \cite{pf03}, these
velocity structures of large vorticity likely trace either
fossil shocks which have generated long-lived vorticity, or
genuine coherent\footnote{so-called because their lifetime
is significantly longer than their period} vortices, both
manifestations of the intermittency of turbulence
dissipation, (\cite{pf00}).

Interestingly, these structures might be the long-searched
sites of turbulence dissipation in molecular clouds, as
proposed by \cite{J98} (hereafter JFPF), and thus the sites
of the elusive warm chemistry put forward to explain the
large abundances of \CHp\
(e.g. \cite{crane95,gredel02,gredel97}), \HCOp\
(\cite{ll96,luli00}) and now \wat\
(\cite{neufeld02,plume04}) observed in diffuse molecular
clouds.  Since all the above observations are absorption
measurements, in the submillimeter, radio and visible
domains, they suffer from line of sight averaging restricted
to specific directions and from a lack of spatial
information regarding the actual drivers of this "warm
chemistry".

The present work is the first attempt at detecting the
chemical signatures of locations identified in nearby clouds
as the possible sites of the "warm chemistry" in diffuse
molecular clouds, \ie\ the locus of enhanced dissipation of
supersonic turbulence. It is part of a series of papers
dedicated to the properties of this new kind of filaments in
dense core environments, in the broad perspective of
understanding how non-thermal turbulent energy is eventually
converted and lost as dense filaments and cores form, and
the timescales and spatial scales involved in the non-linear
thermal and chemical evolution of the dissipative structures
of turbulence.

Unfortunately, dissipative structures amount to very little
mass \ie\ the impulsive releases of energy affect only a
small fraction of the gas at any time and they are difficult
to characterize because their emission is weak in most
tracers.  A multi-transition analysis of CO observations, up
to the $J=4-3$ transition, provides for this gas a full
range of temperatures from 20K up to 200K and densities
ranging from 10$^4$ \cc\ down to 100 \cc,
(\cite{fp96}). Recent observations in the two lowest
rotational lines of \twCO\ and \thCO\ show that the gas
there is associated spatially, although not coincident, with
gas optically thin in the \twCO(1-0) line (\cite{phb04},
\cite{hf06}, hereafter Paper II).

The present  paper reports observations and  analysis of the
\HCOp(1-0)  emission  of a  few  such structures  previously
identified  in  nearby molecular  clouds.  Each  of them  is
characterized  by a  different value  of the  local velocity
shear  measured  in the  \twCO\  lines  in  order to  sample
different strengths (or  stages) of the dissipation process.
In a companion paper, we  present the substructure in one of
these filaments as detected in  a mosaic of 13 fields mapped
in   the  \twCO(1-0)   line   with  the   Plateau  de   Bure
Interferometer (PdBI) (\cite{fph}, herafter Paper III).  The
target fields and the  observations are described in Section
2.  We  detail the  line analysis  done to  derive molecular
abundances in  Section 3.  In Section 4,  we show  that they
cannot  be reconciled  with current  models  of steady-state
chemistry   and   that   an  alternative   model   involving
non-equilibrium heating  and chemistry reasonably reproduces
the main features of our  data, although this model is by no
means final.  The discussion in  Section 5 is devoted to the
limitations  of  our  study  and  the  implications  of  the
interpretation we propose.

\section{Observations}

\subsection{The target fields}

The selected targets belong to the environments of the dense
cores studied by \cite{f98} and \cite{pf03}.  These cores
were selected because of their particularly transparent
environment which helps isolating the gas directly connected
to the cores and minimizes confusion with unrelated
components. One is located in the Polaris Flare, a high
latitude cloud (\cite{heith02}), and is embedded in an
environment where the visual extinction derived from star
counts at high angular resolution ($<1$ arcmin) is $0.6
<A_V<0.8$ mag \cite{camb01}. The other, L1512
(\cite{lee01}), is located at the edge of the
Taurus-Auriga-Perseus complex in a region of low average
column density at the parsec scale ($N_{\HH}$ of a few
$10^{20}$ \cq, as deduced from the low angular resolution
\twCO\Jone\ observations of \cite{ut87}.  L1512 appears to
be a very young dense core because it is one of the least
centrally condensed among nearby dense cores mapped in the
submillimeter emission (\cite{wt94}) and has no signpost of
infall motions toward a central object, such as the
so-called blue line asymmetry (\cite{greg00}).  It is also
one of the denses core where the observed molecular
linewidths are the narrowest, yet not purely thermal
(\cite{fuller}).

\begin{figure*}[t]
  \begin{center}
	\def\wa{0.5\hsize}
	\includegraphics[width=\wa,angle=-90]{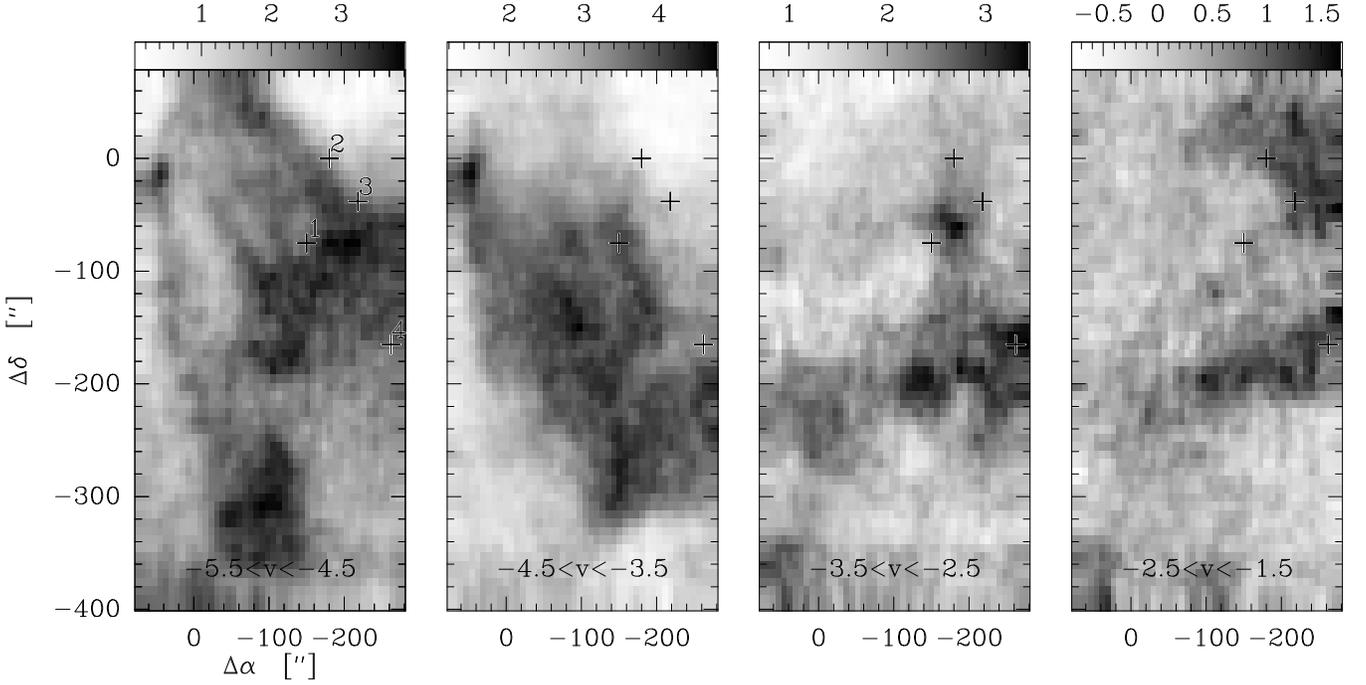}
	\caption{\twCO\Jtwo\ channel maps of the Polaris field.
	  The 4 positions observed in \HCOp\ lie at offsets
	  (-150", -75"), (-180", 0), (-218", -38"), (-262",
	  -165") relative to (0,0): $l_{II}=123.68$,
	  $b_{II}=24.93$.  They are indicated with crosses
	  (positions 1 through 4 respectively). The dense core,
	  not visible in the \twCO\ lines, is located along the
	  eastern edge of the map, south of offset (0,0). The
	  velocity intervals in \kms\ are given at the bottom of
	  each panel.}
	\label{fig:chanpol}
  \end{center}
\end{figure*}

\begin{figure*}[t]
  \def\wa{0.4\hsize}
  \includegraphics[width=\wa,angle=-90]{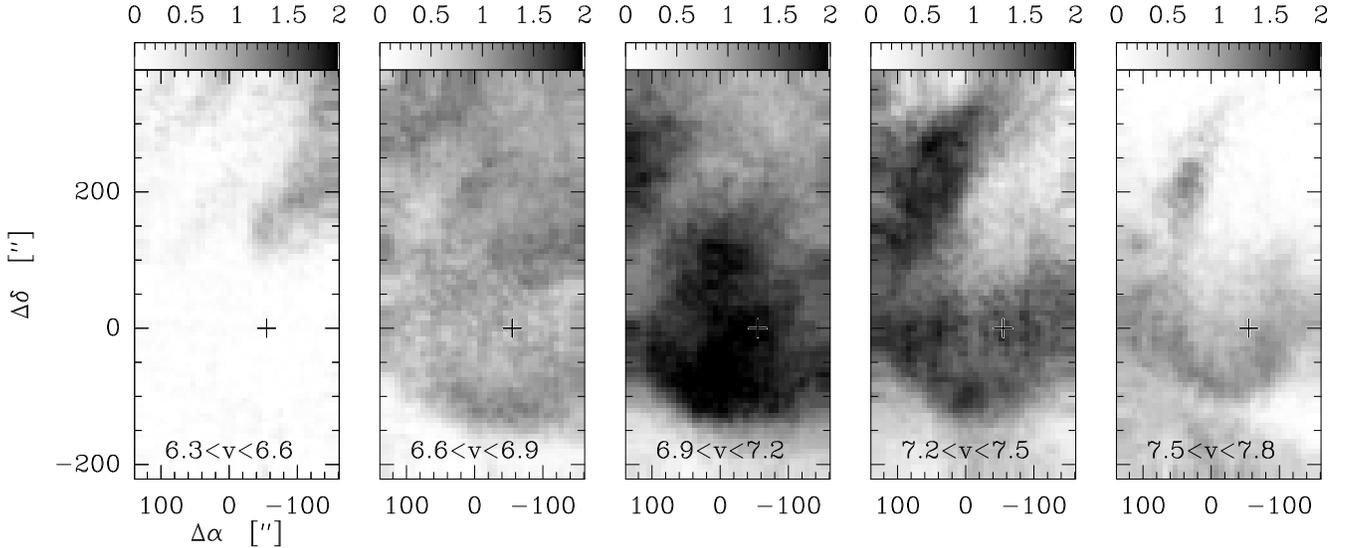}
  \caption{\twCO\Jtwo\ channel maps of the L1512 field. The
	position observed in \HCOp\ is at offset (-55",0)
	relative to (0,0): RA=05h00m54.5s, Dec=32$^\deg$ 39'
	(1950). The dense core is centered on the (0,0) position
	and is about 2 arcmin in diameter (\cite{lee01},
	\cite{pf03}).  The velocity intervals in \kms\ are given
	at the bottom of each panel.}
  \label{fig:chantau}
\end{figure*}

\begin{table*}
  \begin{center}
  \caption{\HCOp\Jone, \thCO\Jone\ and \CtH(3/2-1/2) line
	properties derived from Gaussian decompositions. In
	several cases, velocity components with similar centroid
	and width appear in the \HCOp\ and \thCO\ spectra. They
	are labelled A, B, C, D or E depending on the strength
	of the lines. These are the five cases for which the
	analysis can be completed. The rms noise level $\sigma$
	of the spectra is also given. The value of the centroid
	velocity increment (CVI) measured over a lag of 0.02 pc
	on the \twCO(1-0) line at each position is also given. }
  \begin{tabular}{lllllllllllll}
	\noalign{\smallskip}
	\hline
	\noalign{\smallskip}
 & & & & & \multicolumn{4}{c}{\HCOp\Jone} & \multicolumn{4}{c}{\thCO\Jone}   \\
	\cline{6-9} \cline{10-13}
 & \multicolumn{2}{c}{offsets}& CVI & & $v_i$& $\Tas_i$ & $\sigma$ & $\Delta	v_i$ & $v_i$ & $\Tas_i$ & $\sigma$ & $\Delta v_i$  \\
	& " & " &\kms\ & & \kms\ & K & mK & \kms\ & \kms\ & K & mK  & \kms\   \\
	\noalign{\smallskip}
	\hline
	\noalign{\smallskip}
    Polaris \#1 &-150 &-75 &0.08 &D &-4.7  &0.15 & 5&0.65&-4.7 &2. &16 &0.4 \\ 
                &     &    & &D &-4.3  &0.12 &  &0.6 &-4.3 &2.2& &0.5 \\ 
                &     &    & &C &-3.6  &0.02 &  &0.4 &-3.8 &0.8& &0.3 \\ 
                &     &    & &A &-2.7  &0.015&  &1.1 &-3.4 &0.14& &1.1 \\ 
    Polaris \#2 &-180 & 0  &0.4&C &-4.8  &0.06 &3 &0.4 &-4.6 & 0.6&11&0.5\\ 
                &     &    & & &-4.55 &0.08 &  &0.5 &-3.3 &0.2 & &0.8   \\ 
                &     &    & &B &-3.9  &0.02 &  &1.2 &-2.9  &0.2 & &0.8  \\ 
                &     &    & &A &-2.9  &0.01 &  &2.0 &-3.0 &0.07 & &2.1    \\ 
    Polaris \#3 &-218 &-38 &0.4&C&-4.7 &0.075&3 &0.7 &-4.7 &1.5 &6 &0.5  \\ 
                &     &    & &B &-4.0  &0.03 &  &1.1 &-3.4 &0.3 & & 1.2  \\ 
                &     &    & &A &-2.5  &0.01 &  &1.8 &-1.9 &0.09 & &0.9  \\ 
    Polaris \#4 &-262 &-165&0.1 &C&-4.6  &0.05 &8 &1.0 &-4.5 &0.7 &24 &1.2  \\ 
                &     &    & &C &-3.6  &0.06 & &1.3  &-3.9 &1.4 &   &1.2  \\ 
                &     &    & &C &-3.2  &0.07 & & 0.4 &-3.2 &0.9 &   &0.5  \\ 
                &     &    & & & & &  &  &-2.8 &0.08 &   &2.4  \\ 
	L1512   &-55  & 0  &0.06 &E &7.0  &0.5 &6 &0.3 & 7.0 & 3.7 &50 &0.2 \\ 
                &     &    & &dc &7.2  &1.7 & &0.15 & 7.2 & 4.4 &   &0.3   \\ 
                &     &    & &C &7.2  & 0.05& &0.9  & 7.4  &1.3   &   &0.6  \\ 
	\noalign{\smallskip}
	\hline
 & & & & \multicolumn{4}{c}{\CtH(3/2,2-1/2,1)} & \multicolumn{4}{c} {}  \\
	\cline{5-8} \cline{9-12}
 & \multicolumn{2}{c}{offsets}& & $v_i$& $\Tas_i$ & $\sigma$ & $\Delta	v_i$ &  &  & &   \\
	& " & " & & \kms\ & mK & mK & \kms\ &  & &  &   \\
	\noalign{\smallskip}
	\hline
	\noalign{\smallskip}
    Polaris \#2 &-180 &0  &  &-3.5  &7. &  3.& 3.0&  & & &   \\
        \hline 
  \end{tabular}
  \label{tab:data}
  \end{center}
\end{table*}

Four positions have been observed in the Polaris environment
and one in that of L1512.  They are shown in the \twCO\Jtwo\
channel maps of Figs~\ref{fig:chanpol} and
\ref{fig:chantau}, from the IRAM-30m data of \cite{pf03}.
These positions sample the full range of small scale
velocity shears, following the statistics of the velocity
field of \cite{pf03}. The observed quantity used in such
statistics is the \twCO\ line centroid velocity increment
(CVI) measured for a given lag. The CVI values measured for
a lag of 0.02 pc are given in Table~\ref{tab:data} for each
position.  They correspond to local velocity shears (in the
plane of the sky) ranging from 3 to 20 \kms\ pc$^{-1}$, the
smallest value in the Polaris field belonging to the
Gaussian core of the probability distribution of the CVIs in
that field, while the largest values (positions \#2 and \#3
in Polaris) belong to its non-Gaussian tails. The CVI value
in L1512 also belong to the non-Gaussian tail of the CVI
probability distribution, but is close to the limit of its
Gaussian core.

\subsection{Observations and results}

\begin{figure*}
  \def\wa{0.5\hsize}
  \begin{center}
	\includegraphics[height=\wa,angle=-90]{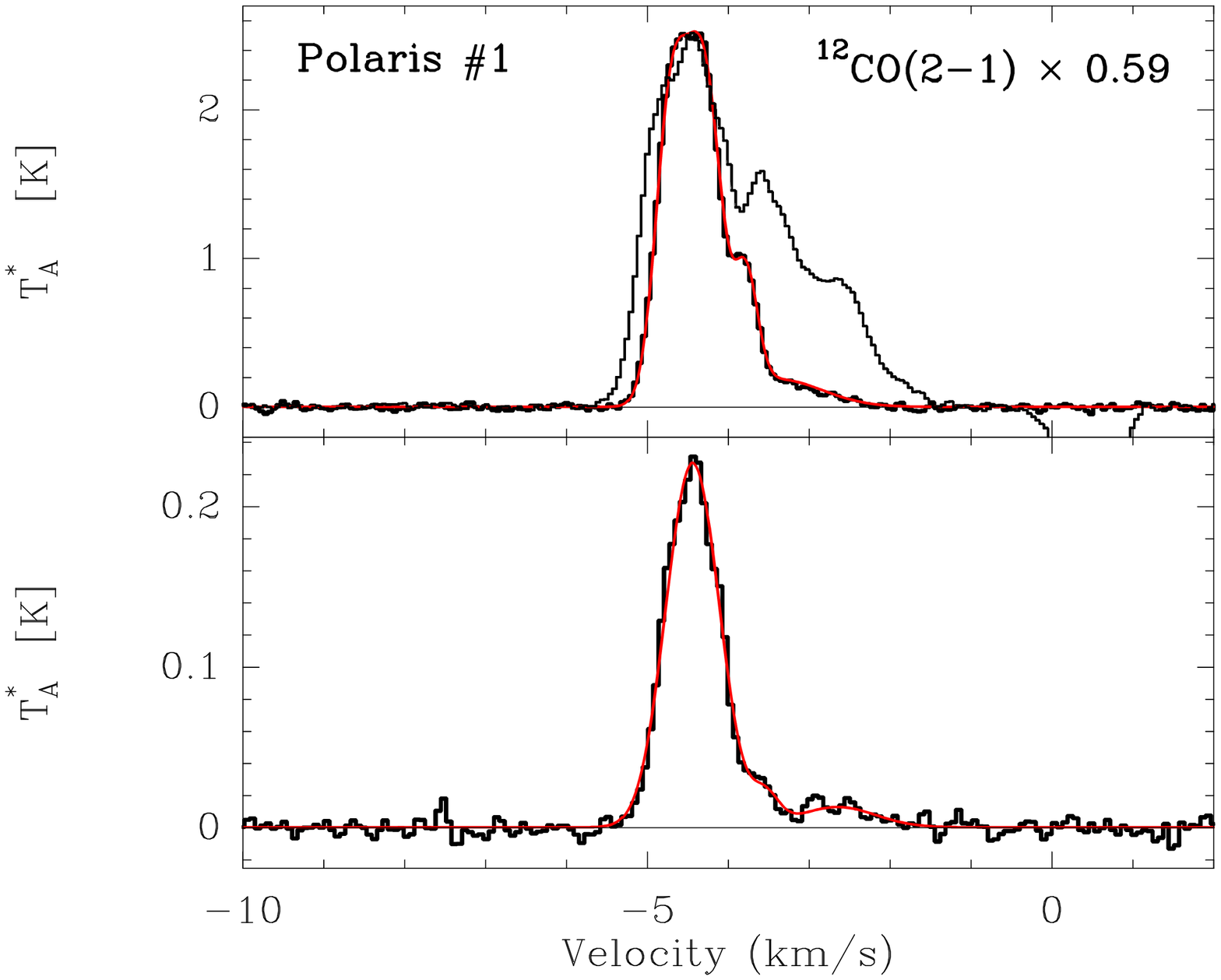}\hfill%
	\includegraphics[height=\wa,angle=-90]{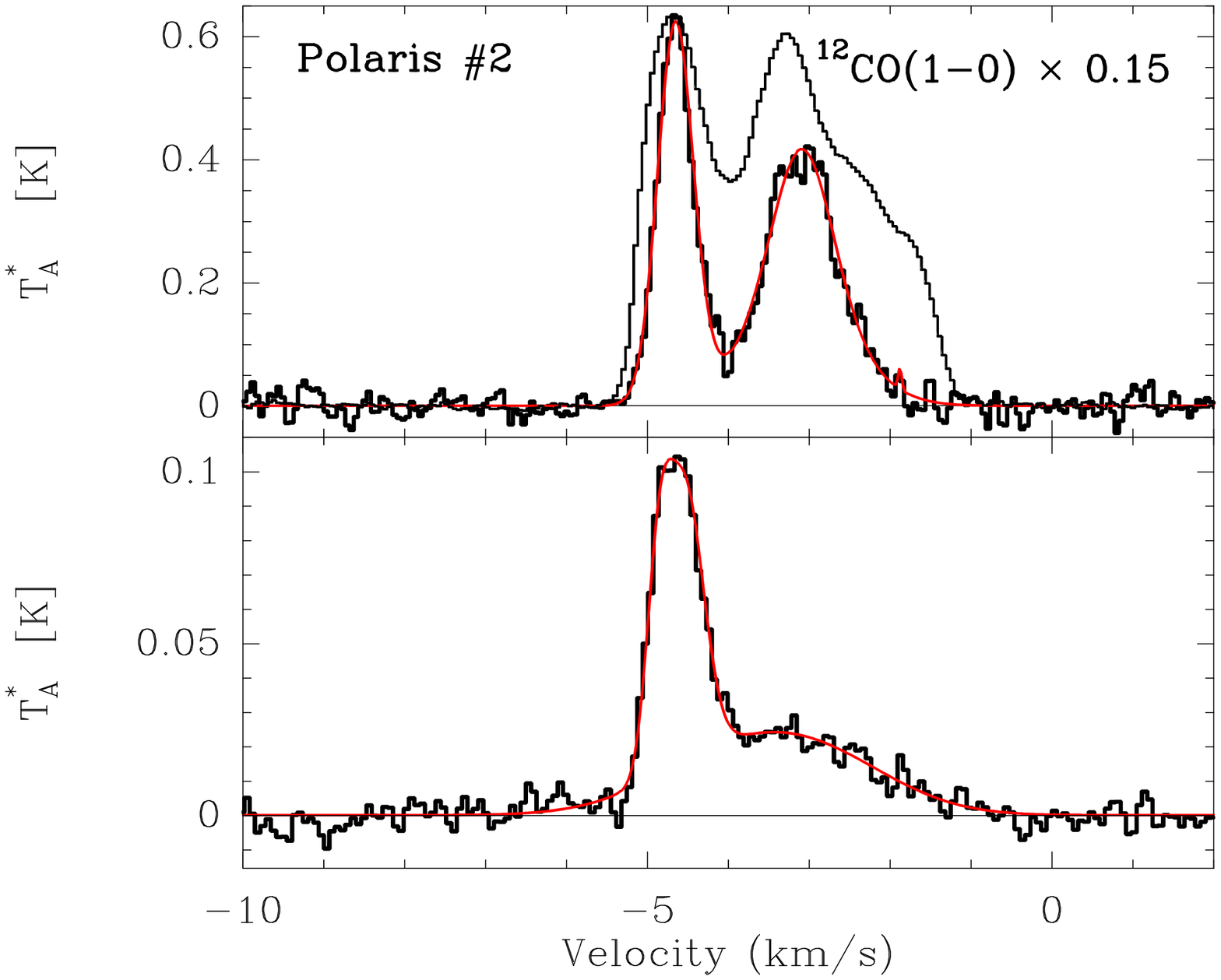}
	\caption{In each of the 4 panels, corresponding to the 4
	  positions observed in the Polaris field, three lines
	  are displayed: \thCO\Jone\ (thick line) and a
	  multi-component Gaussian fit (thick red curve), \twCO\
	  (either \Jone\ or \Jtwo, as indicated) multiplied to
	  fit the \thCO\ temperature scale, and the \HCOp\Jone\
	  at the bottom with a multi-component Gaussian fit
	  (thick red curve).}
	\label{fig:specpol12}
  \end{center}
\end{figure*}

\begin{figure*}
	\def\wa{0.5\hsize}
  \begin{center}
	\includegraphics[height=\wa,angle=-90]{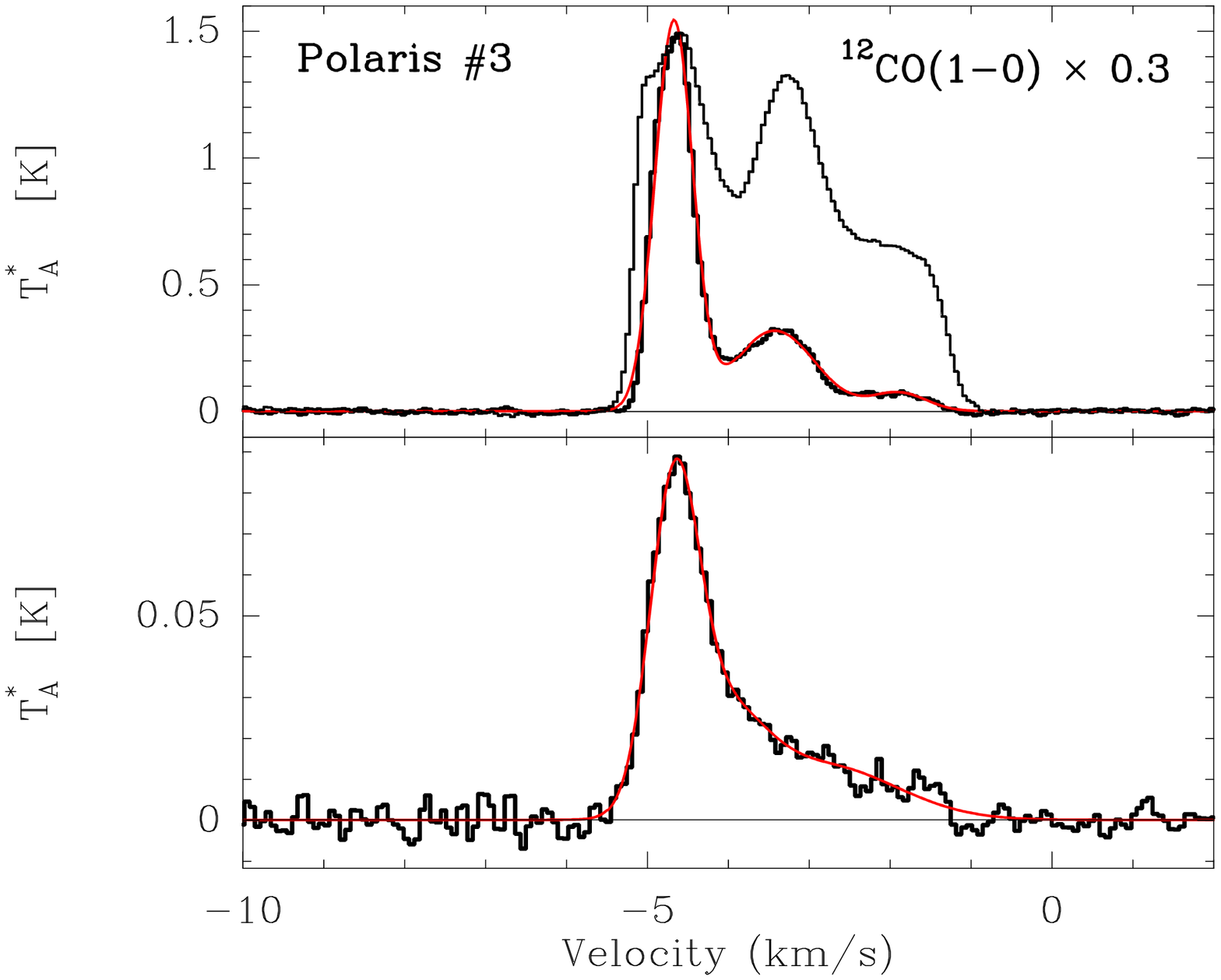}\hfill%
	\includegraphics[height=\wa,angle=-90]{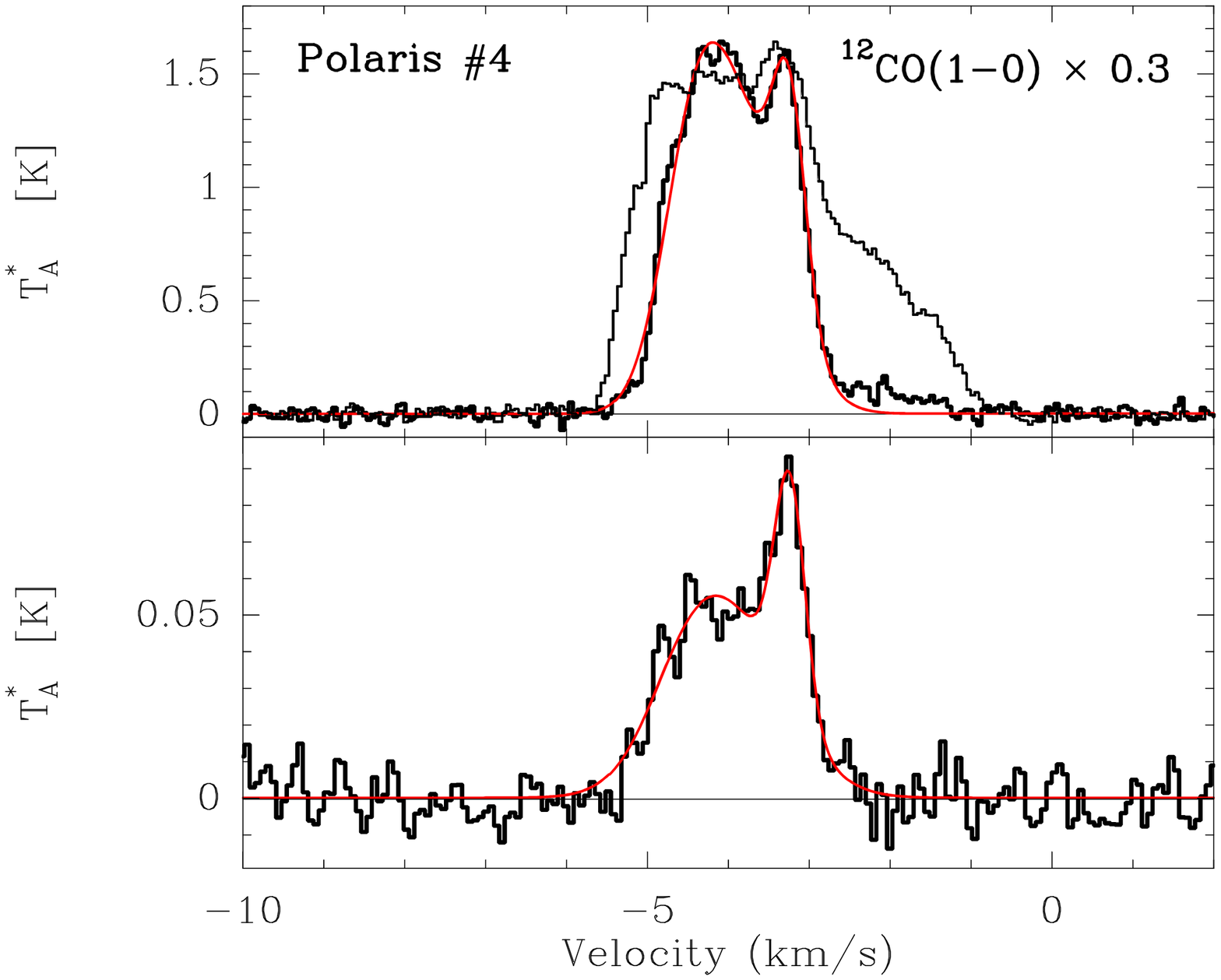}  
	\caption{Same as Fig.~\ref{fig:specpol12} for positions
	  3 and 4 in Polaris.}
	\label{fig:specpol34}
  \end{center}
\end{figure*}

\begin{figure}
  \begin{center}
	\includegraphics[height=0.9\hsize,angle=-90]{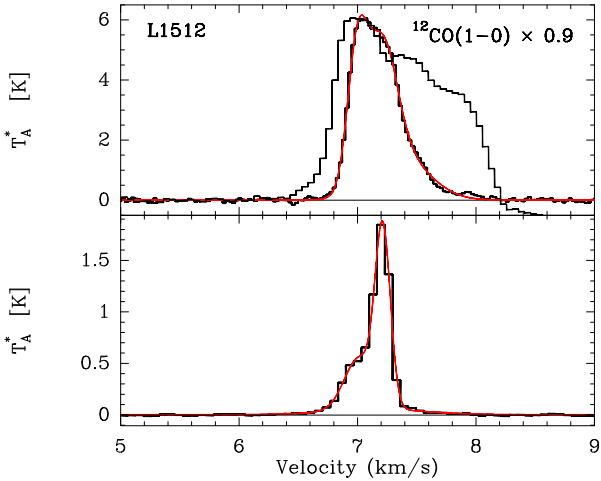}
	\caption{Same as Fig.~\ref{fig:specpol12} for the
	  position in L1512. Note the very different antenna
	  temperatures, with respect to the Polaris field, for
	  the \HCOp\ line in particular, due to the fact that
	  the dense core is sampled by the beam in L1512. The
	  velocity scales for the two fields are also very
	  different.}
	\label{fig:spectau}
  \end{center}
\end{figure}

\begin{figure}
  \begin{center}
	\includegraphics[height=0.9\hsize,angle=-90]{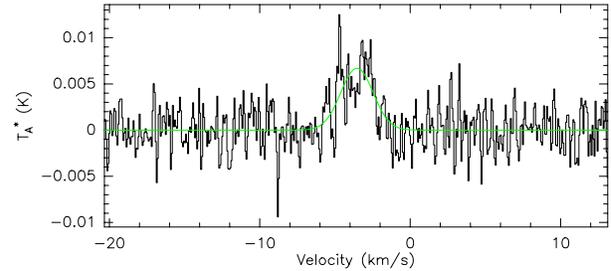}
	\caption{\CtH\ spectrum at position 2 in Polaris.The
      line parameters derived from the Gaussian fit are:
      $v=-3.5$\kms, \Tas=7mK, $\sigma$=3mK and $\Delta v=$3
      \kms).}
	\label{fig:c2h}
  \end{center}
\end{figure}


The observations were carried out at the IRAM-30m telescope
and spread over three summer sessions (1996, 1998 and
1999). A few additional \twCO\ and \thCO\Jtwo\ lines have
been obtained in the framework of another project during a
winter session in 1999.  The weather conditions never
allowed us to observe at high frequency during the summers
and the lines observed were
\HCOp\Jone, \thCO\Jone\ and C$_2$H(3/2-1/2).  Frequency
switching by 7.8 MHz was used to subtract the off line
emission. The autocorrelator had a resolution of 20 kHz
providing a velocity resolution of 0.08 km s$^{-1}$ at 89
GHz. The single sideband noise temperatures of the receivers
greatly improved in 1999 when a new generation of receivers
was installed. At 3mm, the SSB receiver temperatures were in
the range 80-130 K depending on the frequency before 1999
and 50-65 K in 1999. The system temperatures were in the
range 200-250 K before 1999 and between 150 and 250 K in
1999.  The sideband rejections were estimated to be about 20
dB.  At 230GHz, the SSB receiver temperature was about 90 K
with $\approx$ 13 dB gain rejection.  The \HCOp\Jone,
\thCO\Jone\ and \twCO\Jtwo\ line profiles are shown in
Figs.~\ref{fig:specpol12}-~\ref{fig:c2h}
for the five positions, as well as a unique \CtH\ spectrum.

In L1512, the line of sight crosses the dense core, so the
spectra comprise the emission of both the dense core (the
narrow component between 7.1 and 7.3 \kms) and that of its
environment extending from 6.4 to 8.2 \kms. This velocity
interval is assigned to the environment of the core on the
basis of the line profile properties and of the large scale
channel maps of \cite{f98} and \cite{fpp01}. In Polaris, the
lines of sight do not intercept the dense core therefore the
observed emission is from the environment exclusively.

The exceptional signal-to-noise ratio of the line profiles reveals a
large variety of complex non-Gaussian line profiles, with several broad and
weak components.
There is a general agreement between the total velocity
extent of the CO and \HCOp\ lines (except for position \# 4
in Polaris) but the lineshapes are all extremely different.
The velocity components visible in the \thCO\ spectra are
most often present in the \HCOp\ spectra at the same
position.  The \HCOp\ and \thCO\ lines have thus been
decomposed into several Gaussians with the goal of ascribing
a \thCO(1-0) counterpart to each of the \HCOp\
components. The decomposition of these profiles into
Gaussians is not straightforward, and certainly not unique,
but it is a guidance in the analysis of the spectra.  Yet,
as shown in Table~\ref{tab:data}, not all the components
found in the Gaussian fitting procedure of the \HCOp\ lines
have been found successfully in that of the \thCO\ lines and
vice versa. In most cases, the broad \HCOp\ weak emission
has been fitted with a single Gaussian, although in a few
cases, several narrower Gaussians would also fit to the
profile.

\begin{table*}
  \caption{LVG results for the five template velocity
	components. {\bf $^{13}$CO(1-0) line:} range of input
	values for $R$(2-1/1-0) (col. 2) and \Tas(1-0) (col. 3),
	result for the \thCO\ column density per unit velocity
	(col. 4).  {\bf HCO$^+$(1-0) line:} input \Tas (col. 5),
	lower and upper values for the \HCOp\ column density met
	along the curves shown in Fig.~\ref{fig:lvg} (col. 6),
	and corresponding range of \thCO/\HCOp\ ratios
	(col. 7).}
  \begin{center}
  \begin{tabular}{lllllll}
	\noalign{\smallskip}
	\hline
	\noalign{\smallskip}
 &$R$(2-1/1-0) & \Tas(\thCO)  & $N(\thCO)/\Delta v$ & \Tas(\HCOp)
 & $N(\HCOp)/\Delta v$ & $N(\thCO)/N(\HCOp)$  \\
 & &K& \cq/\kms\  & K & $10^{12}$\cq/\kms\  &    \\
	\noalign{\smallskip}
	\hline
	\noalign{\smallskip}
A & 0.3 -- 0.6&0.07 -- 0.2&1.2$\times 10^{14}$ &0.01& 0.1 -- 0.3&1200 -- 400 \\
B & 0.3 -- 0.6&0.2 -- 0.5&3$\times 10^{14}$&0.025 &0.1 -- 0.25 & 3000 -- 1200\\
C & 0.3 -- 0.6& 0.7 -- 1.5 &1.2$\times 10^{15}$&0.09  &0.7 -- 1 &1700 -- 1200\\
D & 0.5 -- 0.6  & 2. -- 3. & 2.$\times 10^{15}$&0.24 &1.2 -- 2.3 &1667 -- 870\\
E &0.55 -- 0.65 & 5. -- 7.   &7.2$\times 10^{15}$&0.5&2.5 -- 4.0&2880 -- 1800\\
	\noalign{\smallskip}
	\hline
  \end{tabular}
  \label{tab:lvg}
  \end{center}
\end{table*}

\begin{figure}
  \begin{center}
	\includegraphics[height=0.9\hsize,angle=-90]{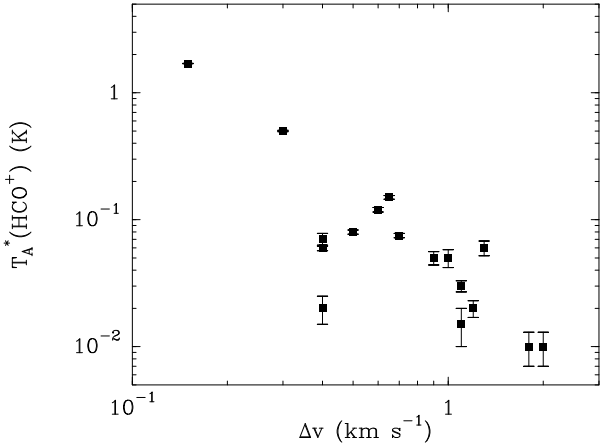}
	\caption{Peak antenna temperature versus half-power
	  linewidth of all the Gaussian components identified in
	  the \HCOp(1-0) line profiles and listed in
	  Table~\ref{tab:data}. }
	\label{fig:idv}
  \end{center}
\end{figure}

We have restricted our analysis of the \HCOp\ lines to the
components which are visible in the \thCO(1-0) spectra \ie\
those for which it exists a \thCO\ counterpart with
comparable centroid velocity and width. They are labelled
with letters in Table~\ref{tab:data}. Instead of analysing
all the profiles separately, we give outlines of the
solutions by splitting the range of line temperatures into
five cases (labelled by letters A to E). Each case is
identified by its \HCOp(1-0) temperature within a factor of
the order of 3, given in Table~\ref{tab:lvg}. Note that in
all cases but E, the \HCOp/\thCO\ line temperature ratio is
$\approx$ 0.1.  The weakest cases are the A's while the E
case approaches the conditions within dense cores (noted
"dc" in Table~\ref{tab:data}).  The characteristics of all
the components are reported in Table~\ref{tab:data}.
Table~\ref{tab:data} shows that the \HCOp\ line temperatures
range between 0.01 and 0.5 K, a range similar to that found
towards $\zeta$~Oph by \cite{ll94} and in the direction of
several extragalactic sources by \cite{ll96}.  An
interesting rough anticorrelation appears between the \HCOp\
peak intensity and the width of each component where
approximately $\Tas_i(\HCOp) \propto \Delta v_i^{-2}$
(Fig.~\ref{fig:idv}).  It will be discussed in Section 4.5.


\section{Column densities and abundances}
\subsection{\HCOp}

\HCOp\ line intensities have been interpreted in the
framework of the Large Velocity Gradient (LVG)
formalism. The collisional excitation due to electrons has
been included in the code following \cite{bhatt}, with
ionisation degrees $x_e=10^{-4}$ and $x_e=10^{-5}$.  The
HCO$^+$--electron rates calculated by \cite{ft01} have not
been used, since they only calculate the rates up to $J=2$,
while for our calculations higher excitation levels have
been included.  We have verified that, at least for higher
temperatures, the rates from \cite{ft01} and \cite{bhatt} do
not differ significantly.  Collisional excitation of \HCOp\
by \HH\ at high temperature is implemented by using the
cross-sections computed by \cite{flower00}.  The excitation
temperature, optical depth and emergent intensity in the
J=1-0 transition of \HCOp\ have been computed for \HH\
densities ranging between 30 and $10^5$ \cc\ and kinetic
temperatures $T_k=10$ and 250 K.  Temperatures as large as
250 K have been considered because observations of the
\twCO\Jthr\ and \Jfo\ transitions of a few positions in the
environment of L1512 reveal \twCO\ intensities larger in
these transitions than anticipated from the low J lines if
the gas is cold and dense. Diluted ($n_{\HH}=200\cc$) and
warm ($T_k=250$ K) gas is a possible solution (Falgarone,
Pety \& Phillips 2001). An upper limit $T_k=200$ K on the
kinetic temperature is provided by the width of the \HCOp\
velocity components, whose median value is 0.6 \kms.  At
$T_k=200$ K, the thermal velocity of \HCOp\ is only
$v_{th}=0.24$ \kms, and the thermal contribution to the
linewidth $\Delta v_{th}=0.57$ \kms.

\begin{figure}
  \begin{center}
	\includegraphics*[width=0.7\hsize,angle=-90]{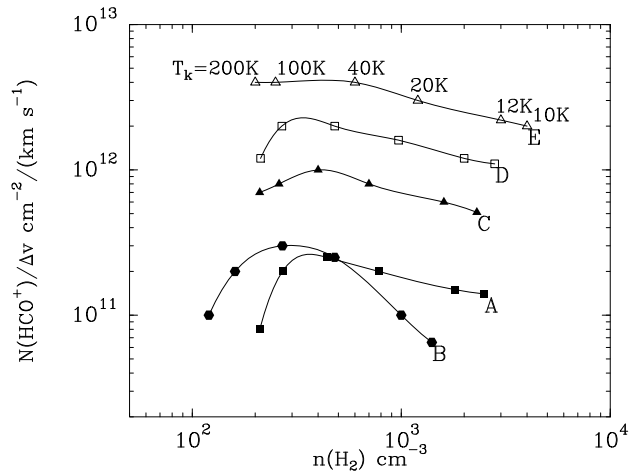}
	\caption{\HCOp\ column densities per unit velocity as a
      function of \HH\ density, for the five template
      components (A to E) for which the LVG analysis is
      possible (see text).  Six values of the kinetic
      temperature are given along the upper curve, the same
      values being marked by the 6 symbols on the other
      curves. Each ($n_{\HH},T_k$) set belongs to the narrow
      track of solutions allowed by the CO data (Sect. 3).}
  \label{fig:lvg}
  \end{center}
\end{figure}

Although weak, the \HCOp\ lines are not optically thin
throughout the intensity range 0.01-0.5 K. For densities
lower than a few 10$^3$ \cc, the optical depth approaches
unity, as the excitation temperature drops towards the
Cosmic Background temperature.  To break the degeneracy of
the problem, we use the \thCO\Jone\ and J=2-1 lines to
constrain the \HH\ density and temperature domain of the
solutions and, as said above, we restrict our analysis to
those components of similar kinematic characteristics in the
\HCOp\ and \thCO\ spectra.  The \thCO\ lines are optically
thin (or almost thin, in the case of L1512) so that the
column density per velocity unit is well determined and the
two constraints of the \thCO\Jone\ antenna temperature and
$R$(2-1/1-0) line ratio define a narrow track of solutions
in the density-temperature plane, which nevertheless spans
the whole range from cold/dense to warm/diffuse gas. Each
[\HCOp, \thCO] pair of components (\ie\ each letter in
Col. 4 of Table~\ref{tab:data}) has then to be interpreted
with the set of values ($n_{\HH}, T_k$) allowed by the
\thCO\ lines.  This method provides determinations of the
\HCOp\ column density per unit velocity $N(\HCOp)/\Delta v$
that do not suffer from the lack of knowledge of the
density, a critical parameter for a line so difficult to
excite.  The results for each ($n_{\HH}, T_k$) set are
displayed in Fig.~\ref{fig:lvg}, where the labels on the
curves indicate the gas kinetic temperature, given only once
on the upper curve. The tracks appear on that figure, from
\HH\ densities in the range 1--2$\times 10^3$~\cc\ at
temperatures below 15~K and to densities between 100 and 300
\cc\ at temperatures above 100~K. In the following
discussion, we restrict the domain of $T_k$ to a range 15 --
20~K to 200~K, the lower value being the temperature of a
gas of density 1 -- 2$\times 10^3$~\cc, shielded from the
solar neighbourhood UV field by 0.5 to 1~mag (\cite{lp04}),
the upper limit corresponding to a fully thermal
contribution to \HCOp\ line widths of 0.57 \kms.

\begin{figure}
  \begin{center}
	\includegraphics[width=0.7\hsize,angle=-90]{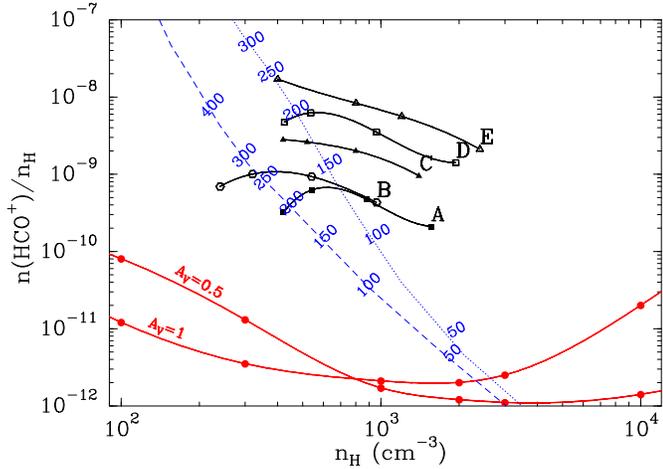}
	\caption{\HCOp\ abundances as a function of the hydrogen
      nuclei density. The five upper curves (thick solid
      curves) are the tracks derived from the observations
      displayed with the same symbols as in
      Fig.~\ref{fig:lvg}.  The kinetic temperature varies
      along each curve and the symbols are located at
      $T_k=$20, 40, 100 and 200 K from right to left. The
      ($n_{\rm H},T_k$) sets are those inferred from the CO
      data. Thick lower curves (red): Results of
      steady-state chemistry models, computed for two
      different values of the UV shielding, $A_v=0.5$ and 1
      mag. Dashed and dotted curves (blue): Tracks of
      non-equilibrium chemical computations along two
      isobaric cooling sequences (see Section 4.4). The
      initial conditions are the same ($n_{\rm H} = 100$
      \cc) except for the shielding from the ISRF: $A_V=0.2$
      mag (dashed) and $A_V=1$ mag (dotted). On each of
      them, the gas temperature decreases from top to bottom
      and is given along each track to allow a comparison of
      the LVG and chemical solutions in the
      [$X(\HCOp),n_{\rm H},T_k$] space.  }
  \label{fig:cool} 
  \end{center}
\end{figure}

In spite of the large range of kinetic temperatures
considered, $20<T_k<200$~K, Fig.~\ref{fig:lvg} shows that
the \HCOp\ column density of each component is determined to
within a factor of a few (see also Table~\ref{tab:lvg}).
Their values for the ensemble of the five template
components span more than one order of magnitude.

The main difficulty in the derivation of the \HCOp\
abundances from the above set of column densities, is the
estimate of the relevant hydrogen column density of each
component.  This is so because each component is defined
only by its velocity and velocity width and contributes to
only a small fraction of the total column density. In the
following expression of the \HCOp\ abundance:
\begin{equation}
  X(\HCOp)= {\left(N(\HCOp) \over \Delta v \right) } \Delta v {1 \over
	n_{\rm H}\, L}
\end{equation}
$N(\HCOp) \over \Delta v$ and $n_{\rm H}$ are derived from
the above LVG analysis for each case, $\Delta v$ is known
from the data. Here, $n_{\rm H}= 2n_{\HH}$ is the density of
protons in a gas where the density of atomic hydrogen has
been neglected, according to PDR models for gas with UV
shieldings above 0.2 mag (\cite{lp04}). The depth $L$ is the
only independent, unknown and free parameter in the
derivation of $X(\HCOp)$.  An estimate of the depth $L$ is
thus needed for each velocity component.  The channel maps
in the \twCO\Jtwo\ lines (Figs.~\ref{fig:chanpol} and
\ref{fig:chantau}) have been used to estimate the spatial
extent in projection, $L$, of each of the velocity
components.  The similarity of the velocity coverages of the
\twCO\ and \HCOp\ lines suggests that at the scale of the
resolution these species are well mixed in space too. The
lengthscale $L$ characteristic of the structures are all of
the order of 0.05 pc in projection.  The actual lengthscale
over which the line forms may be longer by a factor of a
few, less than 10 on statistical grounds (then the filaments
would be the edge-on projection of sheets) or smaller if
substructure exists. In the following, we adopt $L=0.1$~\pc\
for all the components. Note that the estimated total column
density of individual velocity components thus ranges from
$N_{\rm H}=7.5 \times 10^{19}$ \cq\ to ten times more, or
$0.04 < A_V<0.4$, smaller, as expected, than the cloud
visual opacities discussed in Sections 2 and 4.2.  A
velocity coverage $\Delta v= 0.5$ \kms\ has been adopted for
all the components as the median value of the
well-identified components in Table~\ref{tab:data}.  The
broadest components are likely collections of narrower
sub-components.  The results are shown in
Fig.~\ref{fig:cool} for the five cases analysed.  The
symbols are the same as in Fig.~\ref{fig:lvg} and on each
curve they mark the set of kinetic temperatures, $T_k=$20,
40, 100 and 200K from right to left.  According to the
previous discussion, these abundances may have been over or
underestimated by a factor 3 at most. We compare these
determinations to results of chemical models in Section 4.
 
\subsection{\CtH}
The width of the \CtH\ line is large and probably
corresponds to the weakest component seen in the \HCOp\
profile at the same position (case A).  With one transition
detected only, one can simply estimate the column density
assuming the excitation temperature and low opacity.  For
$A_{3/2,2-1/2,1}=4.1\times 10^{-7}$ s$^{-1}$ and an
excitation temperature of 10 K, the inferred column density
is $N(\CtH)=2\times 10^{12}$ \cq. The corresponding
abundance, is $X(\CtH)= 2 \times 10^{-8}$ for an estimated
total column density $N_{\rm H} \approx 10^{20}$ \cq\ from
$n_{\rm H} \approx 500 \cc$ (Fig.~\ref{fig:cool}) and $L =
0.1 $ pc.

At this position, the same component has $N(\HCOp) = 3
\times 10^{11}$ \cq, which makes this component similar to
the weakest lines detected by \cite{luli00} in diffuse
molecular clouds with the PdBI, with an abundance ratio
\CtH/\HCOp $\approx 7$, consistent with the non-linear
relationship they find between these two species.

\section{Comparison with chemical models}

\subsection{Description of the steady-state model for chemistry}
Steady-state \HCOp\ abundances have been computed using the
Meudon PDR code
\footnote{Code available at \\ {\tt
 http://aristote.obspm.fr/MIS/index.html}.}. Models here
 consider semi-infinite clouds of constant density in
 plane-parallel geometry with ionizing photons coming from
 one side. The UV spectral density of the photons is scaled
 according to the Draine radiation field
 (\cite{draine78}). The dust extinction curve is the
 galactic one as parameterized by \cite{fm90}. The
 computation proceeds from the exterior and penetrates into
 the cloud. At each step, chemical and thermal equilibrium
 are reached self-consistently. Radiative cooling makes use
 of the on-the-spot approximation which considers that an
 emitted photon is either re-absorbed locally or escapes
 from the cloud, and line emissivities depend on the
 abundances. Radiative transfer into lines of \HH\ and CO is
 computed using the approximation of \cite{feder79}. The
 depth into the cloud is measured in magnitudes of visual
 extinction (\av). Therefore, $\av=0$~mag at the edge of the
 illuminated cloud, increasing when moving deeper into
 it. One is then free to consider some particular points
 into the cloud, corresponding to particular visual
 extinctions.  We have computed the abundance of \HCOp\ in
 several models with different hydrogen densities, and plot
 the result at two depths in the cloud, $\av=0.5$ and 1~mag
 (see figure~\ref{fig:cool}).  Since the total extinction in
 the direction of the observed positions is smaller than 1
 mag (see Section 2.1) except for the gas in the L1512 dense
 core, not under consideration here, it is reasonable to
 assume that {\it locally} the UV shielding is also less
 than 1 mag, hence the two PDR cases shown in
 Fig.~\ref{fig:cool}.

\subsection{Lack of steady-state chemical solutions}

Fig.~\ref{fig:cool} shows that none of the \HCOp\ abundances
derived from the observations are consistent with
steady-state chemistry in clouds moderately shielded from
the ambient UV field, even at high density.  The
uncertainties in the abundance derivations are large and due
mostly to the unknown extent of the emitting gas along the
line of sight $L$. Note, however, that a depth 10 to 100
times larger (\ie\ 1 to 10 pc) would be required to bring
the curves deduced from the observations down to the domain
of steady-state solutions.  Depths larger than $\sim$ 3pc
are ruled out by the average column density in the fields,
$A_V<1$.  Structures smaller than 0.1 pc are indeed
suggested by the PdBI observations reported in Paper III.
We thus conclude that the observed \HCOp\ abundances cannot
be produced by state-of-the-art models of gas phase
steady-state chemistry in diffuse molecular clouds.

\subsection{Non-equilibrium chemistry}

An alternative scenario relies on the proposal that small
regions bearing chemical signatures different from those of
the bulk of gas are permanently generated in interstellar
matter by bursts of dissipation of turbulence.  The
space-time intermittency of turbulence dissipation is such
that the heating rate due to the local release of
suprathermal energy may exceed those driven by the UV
radiation field. In low density gas and in the UV field of
the Solar Neighborhood, it is the case as soon as
dissipation is concentrated in less than a few \% of the
volume (see the review of \cite{f99}).

Intermittent dissipation locally heats the gas far above its
average equilibrium temperature, over a timescale sufficient
to trigger a specific chemical network driven by endothermic
reactions or involving large activation barriers.  In the
regions heated by non-thermal energy from turbulence, the
endothermic barrier ($\Delta E/k$ =4640 K) of the \Cp\ +
\HH\ $\rightarrow$ \CHp\ + H reaction is overcome and it
occurs at a much higher rate than the slow radiative
associations, \Cp\ + H $\rightarrow$ \CHp\ + $h\nu$ and \Cp\
+ \HH\ $\rightarrow$ \CHtp\ + $h\nu$ (\cite{black78}).  Such
a scenario was first suggested by the good observed
correlation between \CHp\ column densities and rotationally
excited \HH\ (\cite{ld86}).  The \CHp\ ions then initiate
the hydrogenation chain, \CHp\ + \HH\ $\rightarrow$ \CHtp\ +
H and \CHtp\ + \HH\ $\rightarrow$ \CHthp\ + H, which does
not proceed further because \CHthp\ recombines with
electrons faster than it reacts with \HH. In the regions
heated by non-thermal bursts of energy, \CHthp\ thus becomes
the most abundant ion. As discussed in the specific model of
JFPF, it is even more abundant than \CHp\ because its
formation rate stays larger than its destruction rate. It
opens a new route of formation for \HCOp,
\begin{equation}
  \CHthp\ + {\rm O} \rightarrow \HCOp\ + \HH.
\end{equation}   
Large amounts of \HCOp\ are thus expected to be produced by
this chemistry, even in regions weakly shielded from the UV
field. There, \HCOp\ is, in part, a daughter molecule of
\CHp, although other routes are also active (see Section 5).

Following \cite{mko94}, the regions of intermittent
dissipation have been modeled in JFPF by Burgers vortices
threaded by helical magnetic fields, an alternative scenario
to magneto-hydrodynamic (MHD) shocks in the absence of
obvious post-shock compressed layer. The vortex is fed by
large scale motions. Its lifetime is therefore the turnover
time at of these large scales, of the order of ten times (or
more) its own period, $P$. Therefore, the characteristics of
the vortex are not free parameters but imposed by the
ambient turbulence.  Viscous dissipation is concentrated in
the layers of large velocity shear at the edge of the
vortex. In magnetized vortices of the CNM, JFPF show that
decoupling occurs between the neutrals spiralling in and the
ions tightly coupled to steady helical fields.  An
additional local heating rate, due to the friction between
the ions and the neutrals, is thus present.  The chemistry
in the diffuse gas reacts swiftly to the sharp increase of
gas temperature generated by turbulence dissipation.  In the
model of JFPF, relevant for the CNM, the period of the
vortex is $P=600$~yr, its lifetime $\approx 10^4$~yr, and
the minimum time to build up the molecular abundances
specific of the warm chemistry is $t_{warm} \approx 200$~yr.

The sequence of events we consider is therefore: diffuse gas
enters such a vortex (or an ensemble of vortices braided
together), spiralling inward.  It is heated and enriched
chemically during at least 200~yr while it crosses the
layers of largest velocity shear and largest ion-neutral
drift, the {\it active} layers of the vortex.  Then it
enters the central regions of the vortex where the
temperature drops due to the decrease of dissipation. The
gas formerly heated and chemically enriched, starts cooling
down and condensing self-consistently with chemistry. Its
radial velocity there has become vanishingly small.
Eventually, the vortex blows-up, after $\approx 10^4$~yr.

The thermal and chemical evolution of the gas in the {\it
active} warm layers of the vortex has been computed in JFPF.
Here, we follow the time-dependent evolution of the density,
temperature and chemical abundances of the gas, once it has
left these active layers.  An isobaric evolution is assumed.
We use another code with the same chemical network and the
same cooling functions.  The time-dependent evolution is
computed for a given and constant shielding from the UV
field.  The gas temperature is derived from the radiative
cooling rate which depends on the chemical abundances, the
\HH\ density and temperature.

\begin{figure}
  \includegraphics[width=0.95\hsize,angle=0]{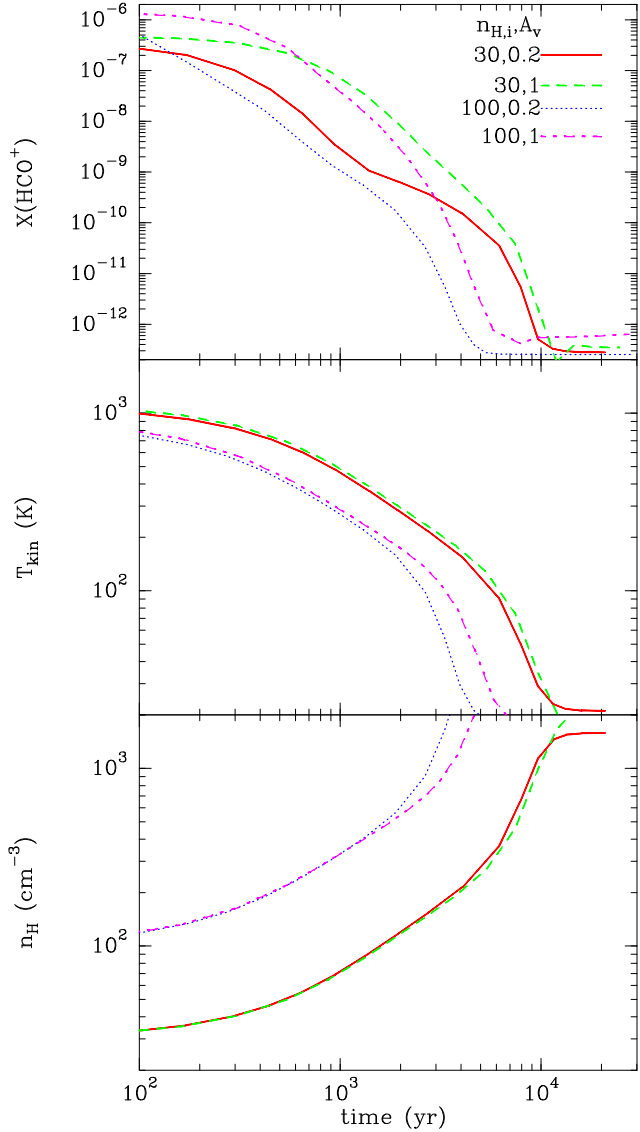}
  \caption{Time-dependent evolution of {\it (top)} the
	\HCOp\ abundance, {\it (middle)} the kinetic temperature
	and {\it (bottom)} the hydrogen density, along the
	isobaric cooling of a parcel of gas formerly enriched
	chemically in a burst of dissipation of turbulent
	energy. The initial densities are $n_{{\rm H},i}$ =30
	(dashed and solid) and 100~\cc\ (dotted and dot-dashed),
	and the shieldings from ISRF are $A_V=$ 0.2 (solid and
	dotted) and 1~mag (dashed and dot-dashed)}
\label{fig:isobar}
\end{figure}

The results are displayed in Fig.~\ref{fig:isobar} for four
different initial conditions in density and shielding from
the UV field.  Two initial densities are considered $n_{\rm
H}=30$ and 100~\cc, representative of the cold neutral
medium (CNM).  The {\it local} heating rate due to turbulent
dissipation, $\Gamma_d$, characterizes the strength of the
burst. Two values have been chosen,
$\Gamma_d=10^{-21}$~\eccs\ for the densest case and
$\Gamma_d=10^{-23}$~\eccs\ for the other, so that the
initial kinetic temperature is close to 10$^3$~K in each
case.  These values correspond to the {\it average} heating
rate due to turbulence dissipation in the CNM, ${\overline
\Gamma_d} = 2 \times 10^{-25} ({\overline n}_{\rm H} /30
\cc)$~\eccs, divided by the estimated volume filling factor
of the gas affected by the burst, or $ \approx 2 \times
10^{-2}$ for the weaker type of burst and $\approx 10^{-3}$
for the stronger and therefore rarer type (see the
discussion in JFPF and \cite{f99}).  Two values of the
shielding from the ambient ISRF are also considered $A_V=$
0.2 and 1~mag to bracket the conditions prevailing in the
gas under study.  The ionisation degree stays close to
$x_e=10^{-4}$ along the isobaric evolution at $A_V=$ 0.2 and
varies between $x_e=10^{-4}$ and $10^{-5}$ in the other
case, following the chemical evolution of the ions. These
upper and lower values of $x_e$ are those used in the LVG
code. The initial chemical abundances are those produced by
the warm chemistry in the active layers of the vortex, as
computed by JFPF.
 
It is remarkable that the signatures of the warm chemistry
are kept by the gas for more than 10$^3$ years after the gas
has escaped the active layers.  The curves of
Fig.~\ref{fig:isobar} illustrate the marked non-linearity of
the chemical evolution: (1) large differences are visible
after only 10$^3$ years between the tracks followed by two
parcels of gas with very similar initial conditions, (2) the
drop of $X(\HCOp)$ is by 6 orders of magnitude while the
density and temperature vary by a factor 50 only.

\subsection{Existence of [$X(\HCOp),n_{\rm H},T_k$] tracks crossing
  the observation domain}

We show below that we may have detected the decaying
signatures of the warm chemistry generated by bursts of
dissipation of turbulence. The time-dependent evolution of
an enriched parcel of gas once the heating due to
dissipation of turbulence has dropped, is a curve in the
[$X(\HCOp),n_{\rm H},T_k$] space which depends on the
initial conditions and UV shielding (see
Fig.~\ref{fig:isobar}). It is therefore not straightforward
to compare it with observational results.  As an
illustration that such an evolution is consistent with our
data, we have drawn two of the tracks of
Fig.~\ref{fig:isobar} (those corresponding to the largest
initial density) in the [$X(\HCOp)$--$n_{\rm H}$] plane of
Fig.~\ref{fig:cool}. On each of them, the gas temperature
decreases from top to bottom from about 500 to 50~K.  In
that plane, the LVG solutions and the cooling tracks are
almost perpendicular. The only temperature range where they
are consistent are over the range 100--300~K: the low
extinction track (dashed curve) is consistent with the LVG
solutions for components A and B and the more shielded
evolution (dotted curve) for all the others (C, D and E).

\begin{table*}
  \caption{Sets of non-equilibrium \HCOp\ abundances
	consistent with \HCOp\Jone\ line observations. Each
	solution is given as a range of values for the set
	[$X(\HCOp), n_{\rm H}, T_k$] corresponding to the range
	of observed values (cases A to E). The parameters of the
	models are the turbulent heating rate $\Gamma_d$, the
	shielding from the UV field $A_v$ and the initial
	density $ n_{{\rm H},i}$.}
  \begin{center}
	\begin{tabular}{llllll}
	  \noalign{\smallskip}
	  \hline
	  \noalign{\smallskip}
	  $\Gamma_d$ & $A_v$ & $ n_{{\rm H},i}$ & $X(\HCOp)$ &$n_{\rm H}$  & $T_k$ \\
	  erg cm$^{-3}$ s$^{-1}$ & & \cc\ &  & \cc\ & K \\
	  \noalign{\smallskip}
	  \hline
	  \noalign{\smallskip}
	  10$^{-21}$ & 0.2 & 100 & 8--2 $\times 10^{-10}$ & 320--500  & 250--150 \\
	  10$^{-21}$ & 1  & 30 &3--0.5$\times 10^{-9}$ &130--200 & 250--170 \\
	  10$^{-21}$ & 1 & 100 & 6--1.6$\times 10^{-10}$ & 700--900  &150--120\\
	  10$^{-23}$ & 0.2 & 30 & 9--3$\times 10^{-9}$ &80--130 & 250--160\\
	  10$^{-23}$ & 1 & 100 & 7--1$\times 10^{-9}$ &100--120 & 250--220\\
	  \noalign{\smallskip}
	  \hline
	\end{tabular}
  \label{tab:absol}
  \end{center}
\end{table*}

The range of values of the \HCOp\ abundances, density,
kinetic temperature of the part of the isobaric cooling
consistent with the observations (cases A to E) are given in
Table~\ref{tab:absol} for five different sets of initial
conditions, including the strength of the dissipation burst
which determines the local heating rate, $\Gamma_d$.

We have not explored a larger domain of initial conditions
because our goal here is only to illustrate the
compatibility of such a non-equilibrium chemistry with the
data. A more detailed analysis requires further constraints
to treat the many non-linearities of the physics and
chemistry involved and explore a relevant part of the vast
parameter space.  Moreover, our model, although complex in
terms of coupled processes, is still too crude.  For
instance, we have neglected the initial dynamical expansion
of the warm gas because the radiative cooling and chemical
timescales, up to $\sim$ a few thousand years, are smaller
than the expansion time of the warm gas under the effect of
pressure gradient, just after the burst of dissipation. The
timescale for a warm structure of initial thickness $L_0$ to
double its size under expansion at the sound velocity in the
ambient cold neutral medium ($c_S \sim 0.5$ \kms) is
2$\times 10^5 (L_0/0.1 \pc)$ yr. Would they be much smaller,
the timescale would be shorter.  The Lorentz force acting on
the ions somewhat delays the expansion at a rate which
depends on the gas ionisation degree and geometry of the
helical magnetic fields.

Nonetheless, the above comparison brings to light a few
interesting results: all the solutions provided by the
non-equilibrium chemistry consistent with the observations
are warmer than 100 K (Table~\ref{tab:absol}) and the
inferred \HCOp\ abundances are all larger than $ 2\times
10^{-10}$ and as large as $\approx 10^{-8}$.  They are more
than one order of magnitude above the values computed at
steady-state at densities less than $10^3$ \cc.  It is
remarkable that these abundances bracket the value
$X(\HCOp)=2 \times 10^{-9}$ determined by \cite{ll96} and
\cite{lilu00} from their observations of the \HCOp\Jone\
line in absorption against extragalactic continuum
sources. These lines of sight sample ordinary interstellar
medium of low visual extinction (many of the extragalactic
background sources are visible QSOs) as opposed to dense
cores and star forming regions. It is thus plausible that
their observations sample the kind of molecular material we
discuss here, their inferred abundances being an average
along the lines of sight of \HCOp-poor and \HCOp-rich
regions. Last, results are weakly sensitive to the strength
of the dissipation burst, as if the actual initial
conditions were rapidly forgotten.

\subsection{The \HCOp\ enrichment and the strength of the velocity shear}

As seen in Table~\ref{tab:data}, the five observed positions
do not sample gas with similar kinematic properties: two
positions (Polaris \# 2 and 3) lie on the locus of largest
velocity shears in the Polaris field (\ie\ populating the
non-Gaussian tails of the probability distributions of CVIs,
\cite{pf03}). The others correspond to smaller velocity
shears.  Against all expectation (Table~\ref{tab:data}), the
largest \HCOp\ abundances (D, E and dc components) are found
at those positions of smallest CVIs, while the set of least
chemically enriched components (A, B and C) are found at the
positions with the largest CVIs.  Since the larger the
velocity shear, the larger the dissipation and heating rates
and the gas temperature, one expects that the chemical
enrichment would increase with the velocity shear or the
CVIs, but it is not what is observed.

These results suggest a broader time sequence, encompassing
that discussed in the previous section, in which the
increase \HCOp\ abundance is a measure of the elapsed
dissipation: the longer the dissipation has been going on,
the smaller the velocity shear (\ie\ the vortex is fading)
and the larger the quantity of chemically enriched gas
produced.  Such a sequence is also suggested by the rough
anticorrelation displayed in Fig.~\ref{fig:idv}. The \HCOp\
column density of each component being close to $\Tas_i
\Delta v_i$, since the line is not very optically thick, it
increases as the velocity dispersion decreases, as $\Delta
v_i^{-1}$.  The observed quantity of chemically enriched gas
is produced at the expense of the non-thermal kinetic
energy, traced by $\Delta v_i$.

\section{Discussion and perspectives}
   
\subsection{The complex space-time average hidden in the observed 
\HCOp\ emission lines.} 

The results of the last section seem to be in contradiction
with the relaxation scenario described above, in which the
\HCOp\ abundance decreases with time. Another, probably
related question, is: Why, in regard of the six orders of
magnitude spanned by the \HCOp\ abundance in the proposed
non-equilibrium evolution, the observations in emission
point to a much narrower range of abundances, $2 \,10^{-10}
<X(\HCOp)<10^{-8}$?

This may be understood in a framework in which, several
phases of the evolution are present simultaneously in the
beam, for individual vortices much thinner than the
beamsize, here 0.015 pc (or $\approx 3000$~AU) while
vortices, in the JFPF model, have a radius $\approx 15$AU.
Observations sample a mixture of gas: the most diluted and
\HCOp-rich, still in the warm active layers of the vortices,
as long as the vortex is alive, gas already out of the warm
active layers and relaxing, up to the densest, \HCOp-poor
gas.  The older the burst, the larger the amount of relaxed
gas.  The signal detected in emission is sensitive to a
combination of the \HCOp\ column density, or simply the
abundance $X(t)$ assuming the total gas column density being
processed is constant with time, the density $n_{\rm H}(t)$
that enters the line excitation and a term involving the
duration of the phase, $\approx t$, given the broad range of
timescales involved.  We have computed the time integral of
$X(t)\,n_{\rm H}(t)$, taken as a template for the signal in
emission, assuming that we observe all the phases in the gas
accumulated during at least $\approx 10^4$~yr, the lifetime
of the vortex. For all the cases, this quantity increases
linearly with time up to a maximum at about $10^3$~yr and
does not increase further.  The observations in emission are
thus likely to be most sensitive to the the relaxation phase
because it lasts longer and involves denser gas than the
brief phase of chemical enrichment in gas too diluted to
significantly excite the \HCOp(1-0) line.

At the opposite, density does not weight the observations in
absorption. The warmest and richest phase of the evolution,
may have been detected by \cite{lilu00} as the broad and
weak absorption seen in the direction of B0355+508 over the
whole velocity extent of the H~I absorption line.

More detailed comparison between abundances inferred from
observations in absorption and in emission involves time and
space averages which cannot be achieved at this stage of the
study. The present observations have been obtained on one
single ensemble of such dissipative structures, possibly
generated simultaneously. Thus, space/time average is likely
more simple than that required to interpret absorption
measurements that sample depths of gas at parsec scales.

\subsection{Further support for non-equilibrium chemistry: "warm" molecules abundance ratios}
The present work stresses the need for a non-equilibrium
mechanism to explain the large abundances of \HCOp\ detected
in gas of moderate density and low shielding from the
ambient ISRF. In this scenario, the gas is driven locally
past a high temperature threshold by an impulsive release of
turbulent energy. Once the gas has escaped the region of
large dissipation, it cools down and condenses and the
chemistry evolves accordingly.

\begin{figure}
  \includegraphics[height=\hsize,angle=-90]{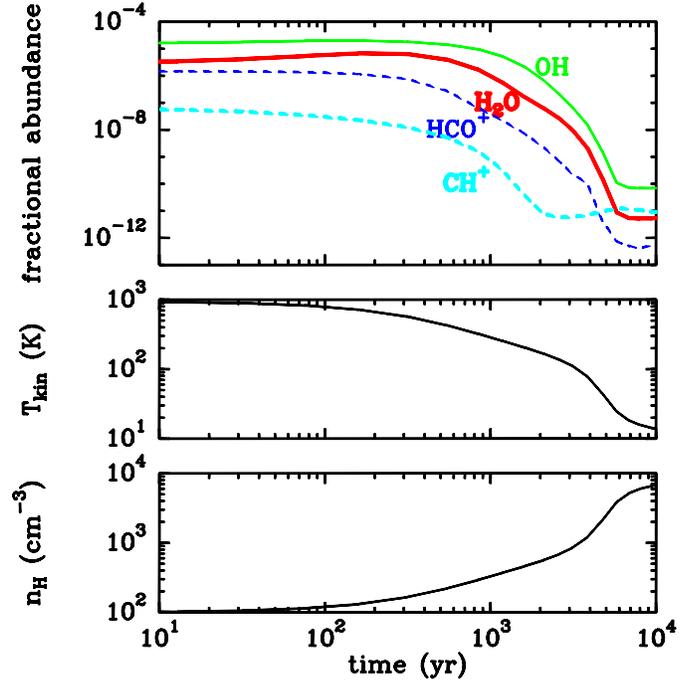}
  \caption{Time-dependent evolution of {\it (top)} the OH
    (thin solid), \wat\ (thick solid), \HCOp\ (dashed), and
    \CHp\ (thin dashed) abundances, {\it (middle)} the
    kinetic temperature and {\it (bottom)} the total
    hydrogen density, along the isobaric cooling of a parcel
    of gas formerly enriched chemically in a burst of
    dissipation of turbulent energy. The initial conditions
    here are $n_{{\rm H},i}$ =100~\cc,
    $\Gamma_d=10^{-21}$~\eccs\ and the shielding from ISRF
    $A_V$=1.}
\label{fig:hotmol}
\end{figure}

\begin{figure}
  \includegraphics[height=\hsize,angle=-90]{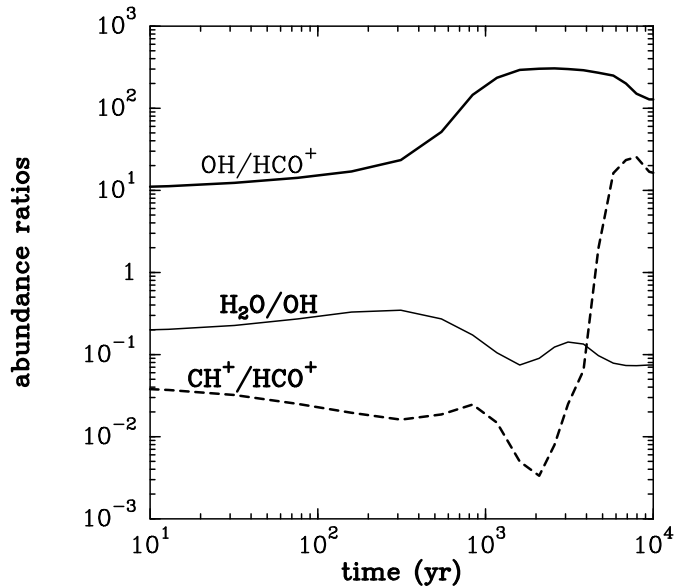}
  \caption{Time-dependent evolution of abundance ratios,
    OH/\HCOp\ (thick), \wat/OH (thin) and \CHp/\HCOp\ (thick
    dashed), along the same isobaric cooling as in
    Fig.~\ref{fig:hotmol}.}
\label{fig:ratio}
\end{figure}

Our results are independent of the details of the processes
at the origin of the warm chemistry. \cite{fp98} and
\cite{gredel02} have shown that slow MHD shocks of
velocities $V_s \sim 10$ \kms\ reproduce the OH/\HCOp\
abundance ratios observed in the diffuse medium
(\cite{ll96}).  In MHD shocks, as in magnetized coherent
vortices, gas is temporarily heated to temperatures as large
as $10^3$ K. The triggered warm chemistry is characterized
by coexisting large abundances of OH, H$_2$O, \HCOp\ and
\CHp, water and \HCOp\ being daughters of OH and \CHp,
respectively (JFPF).  Their time-dependent evolution in the
relaxation phase, is displayed in Fig.~\ref{fig:hotmol} for
one of the cases discussed in Section 4 (initial density
$n_{{\rm H},i}=100$~\cc, $\Gamma_d=10^{-21}$~\eccs\ and
$A_V=1$).  The abundance ratios along this evolution are
interesting in their marked differences
(Fig.~\ref{fig:ratio}).  The \wat/OH ratio varies by less
than a factor 3 along the time-dependent evolution and up to
$10^3$~yr after the beginning of the relaxation, stays close
to the values found in diffuse molecular clouds by the {\it
SWAS} satellite (\cite{neufeld02}, \cite{plume04}).  The
OH/\HCOp\ ratio varies by more than a factor 10 along the
same evolution. We note however that its variations bracket
the average value OH/\HCOp$\sim$ 30 found by \cite{ll96} in
their random sampling of the diffuse ISM provided by their
absorption lines survey towards extragalactic sources.  In
contrast, the \CHp/\HCOp\ ratio varies very little, before
$10^3$ yr only.  The value of the ratio, though, critically
depends on the density (JFPF).  This follows from the fact
that in our model of warm chemistry, \HCOp\ is not only a
daughter molecule of \CHp. It also forms via other routes
such as CO$^+$ + \HH\ $\rightarrow$ \HCOp\ + H, triggered by
the endothermic charge exchange between \Hp\ and O ($\Delta
E/k=227$ K), and \Hthp\ + CO $\rightarrow$ \HCOp\ + \HH,
also slightly enhanced in warm chemistry because both CO and
\Hthp\ have increased abundances in the warm layer of the
vortex.  This route may even be a dominant formation scheme
for \HCOp, if the large abundance of \Hthp\ detected on the
line of sight towards $\zeta$ Per by \cite{mccall} happens
to be the rule in the diffuse ISM.

There is also an efficient channel to form \CtH\ in the warm
chemistry. It proceeds via the endothermic reaction ($\Delta
E/k = 1760$ K) CH + \HH $\rightarrow$ \CHt\ + {\rm H},
followed by a series of fast reactions: \Cp\ + \CHt\
$\rightarrow$ \CtHp\ + {\rm H}, the hydrogenation of \CtHp\
into \CtHtp\, the recombination of \CtHtp\ and terminating
with photodissociation of \CtHt\ which provides \CtH.

These results are encouraging and suggest that impulsive
chemical enrichment of the CNM by turbulence dissipation
bears observable signatures up to several 10$^3$~yr after
the end of the enrichment.

\subsection{Large scale signatures of small scale activity. } 

Other signatures, independent of chemistry, have been found
for the existence in the CNM of gas at a few hundred
Kelvin. They comprise collisional excitation of \HH\ far
from UV sources susceptible to generate fluorescence
(\cite{gry02,lacour05,f05}). In the former cases, \HH\ is
detected in UV absorption against nearby late B stars, in
the third case, the pure rotational lines of \HH\ have been
observed along a long line of sight across the Galaxy
avoiding star forming regions. The rotational temperature of
the warm \HH\ in these data is 276 K.  In all cases the
fraction of warm \HH\ in the total column of gas sampled is
the same, of the order of a few percent.  It is the same
fraction as that required in the Solar Neighbourhood to
reproduce the large observed abundances of \CHp\ within the
CNM (JFPF). A few percent of warm gas, for which UV photons
cannot be the sole heating source, is ubiquitous in the cold
diffuse medium.

Last, we would like to stress the spatial scales presumably
involved in the framework of the existing model, where
individual vortices have transverse hydrogen column
densities as small as $N_{\rm H} \approx 2\times
10^{16}$\cq\ corresponding to initial values $n_{\rm
H,i}=100$\cc\ and $l_i=15$AU (JFPF).  We illustrate the
point with the solutions of case C: $N(\HCOp) \approx
8\times 10^{11}$\cq, $X(\HCOp)\approx 2\times 10^{-9}$, $n_H
\approx 500$\cc. The total hydrogen column density of
observed material is thus $N_{\rm
H}=N(\HCOp)/X(\HCOp)=4\times 10^{20}$\cq, corresponding to
2$\times 10^4$ vortices in the IRAM beam (here we assume
that the isobaric evolution occurs at constant mass, so that
the transverse column density of vortices remains the same).
For the gas at 500\cc, the transverse size has a thickness
of about 3 AU. The ensemble of the 2$\times 10^4$ vortices,
probably braided together, as vortex tubes do e.g
\cite{ppw94}, would form an elongated structure about 450 AU
thick, if the vortices are in contact with one another.  The
IRAM beam at the distance of the sources is a few times
larger and the size of the structure seen with the PdBI is
of this order of magnitude (Paper III).  This rough estimate
just shows that the existence of chemical inhomogeneities at
AU scales is not ruled out in the diffuse ISM.  It would
certainly help understanding the remarkable similarity of
the OH and \HCOp\ absorption line profiles reported by
\cite{lilu00}.

Chemical and thermal relaxation timescales are so short
compared to the dynamical lifetime of molecular clouds (of
at least a million year) that all the stages of the above
evolution should co-exist along a random line of sight.  At
present, the receiver sensitivity and the telescope beam
sizes provide, when observed in emission, a signal which is
the space and time average of myriads of small unresolved
structures, likely caught at different epoch in their
evolution. Molecular line observations in absorption in the
UV, visible and now submm domains are much more sensitive
and have presumably already started to sample this
apparently untractable complexity.

\section{Summary}

This work is part of a broad study dedicated to those
singular regions in the environment of low mass dense cores,
characterized by non-Gaussian velocity shears in the
statistics of the turbulent velocity field of the clouds.
These regions have been proposed to be the sites of
intermittent dissipation of turbulence in molecular clouds,
and the large {\it local} release of non-thermal energy in
the gas there has been shown to be able to trigger a
specific warm chemistry, in particular to explain the large
abundances of \CHp, \wat\ and \HCOp\ observed in diffuse
gas.

We report here the first detection of \HCOp(1-0) line
emission focussed on two of these singular regions.  The
\HCOp\ abundances inferred from the data cannot be explained
by steady-state chemistry driven by UV photons: the gas
temperature is too low and the \HCOp\ abundances formed are
too small by more than an order of magnitude.  As a
follow-up of the model proposed by JFPF, we have computed
the time-dependent evolution of the specific chemical
signatures imprinted to the gas by the short bursts of
turbulence dissipation (a few 100 years), once the gas has
escaped the layers of active warm chemistry. We show that
the signature of the warm chemistry survives in the gas for
more than 10$^3$~yr and that the density (a few 100 \cc),
the temperature (a few hundred Kelvin) and the large \HCOp\
abundances ($2 \times 10^{-10}$ to $10^{-8}$) inferred from
the observations, can be understood in the framework of this
thermal and chemical relaxation, while steady-state
chemistry driven by UV photons fails by more than one order
of magnitude.

The large abundances detected in the present work, and
probably the many signatures of a warm chemistry now found
in diffuse gas, are thus likely due to the slow chemical and
thermal relaxation of the gas following the large impulsive
releases of energy at small scale due to intermittent
dissipation of turbulence.  The detailed confrontation of
our model with available data relevant to the existence of a
warm chemistry ongoing in cold diffuse gas involves complex
time and space averaging, but abundance ratios among \wat,
OH, \HCOp\ and \CHp\ already suggest that such
non-equilibrium processes at AU scale are at work in diffuse
clouds.

\acknowledgement{We are particularly grateful to our
  referee, Harvey Liszt, for his supportive and stimulating
  report who encouraged us to more thoroughly interpret our
  data set and develop the discussion into further detail. }

\end{document}